\documentclass[11pt,oneside,twocolumn]{amsart}
\usepackage[margin=2cm]{geometry}                
\usepackage{graphicx}
\usepackage{amssymb}
\usepackage{epstopdf}
\usepackage[foot]{amsaddr}
\begin{document}

\title{Haptic BCI Paradigm based on Somatosensory Evoked Potential}
\author{Tomasz M. Rutkowski$^{1,2}$, Hiromu Mori$^1$, Yoshihiro Matsumoto$^1$, Zhenyu Cai$^1$,\\ Moonjeong Chang$^1$, Nozomu Nishikawa$^1$, Shoji Makino$^1$, and Koichi Mori$^3$}
\address{$^1$TARA Life Science Center, University of Tsukuba, Japan}
\address{$^2$RIKEN Brain Science Institute, Wako-shi, Japan}
\address{$^3$National Rehabilitation Center for Persons with Disabilities, Tokorozawa, Japan}
\urladdr{http://about.bci-lab.info}
\email{tomek@tara.tsukuba.ac.jp}
\date{July 15, 2012} 

\begin{abstract}
A new concept and an online prototype of haptic BCI paradigm are presented. Our main goal is to develop a new, alternative and low cost paradigm, with open-source hardware and software components. We also report results obtained with the novel dry EEG electrodes based signal acquisition system by \texttt{g.tec}, which further improves experimental comfort.
We address the following points: a novel application of the BCI; a new methodological approach used compared to earlier projects; a new benefit for potential users of a BCI; the approach working online/in real-time; development of a novel stimuli delivery hardware and software. 
The results with five healthy subjects and discussion of future developments conclude this submission.
\end{abstract}

\maketitle
\thispagestyle{empty}

\vspace{-0.7cm}
\section{Introduction}

The state of the art BCI relies mostly on visual and imagery paradigms~\cite{bciBOOKwolpaw}, which require longer training or good vision from the subjects. Many late stage ALS suffering patients have weaker or lost sight. Thus, as the the alternative BCI solutions, the auditory~\cite{tomekHAID2011} or haptic (somatosensory) modalities~\cite{sssrBCI2006,JNEtactileBCI2012} are proposed in order to enhance brain-computer interfacing comfort or to boost information transfer rate (ITR).
A concept of somatosensory (haptic)  modality creates a very interesting possibility to target tactile sensory domain, which is not as demanding as vision during operation of machinery or visual computer applications. A potential somatosensory BCI paradigm is thus less mentally demanding. The first successful trial to utilize responses to somatosensory stimuli utilized steady-state ones~\cite{sssrBCI2006} with lower frequencies exciting mostly Meissner-endings of human finger tips~\cite{tactileNATURE2009}. There is also a very recent development with somatosensory ERP based paradigm using tactile Braille stimulators creating static ``push-stimuli''~\cite{JNEtactileBCI2012}.
We propose to target the higher tactile stimulating frequencies in form of $10$ms long bursts to stimulate Pacini-endigs~\cite{tactileNATURE2009}. Such stimulus creates ``a quick touch sensation'' delivered through tactile exciters as depicted in Figure~\ref{fig:hand}. The concept is perfectly suited for a P300 response based BCI paradigm design. 

\section{Methods}

The report results with the haptic BCI paradigm are based on \texttt{the g.MOBIlab+} portable eight channels EEG amplifier and \texttt{the g.SAHARA} dry electrodes set by \texttt{g.tec}. The eight dry EEG electrodes have been placed at the $Cz$, $CPz$, $POz$, $Pz$, $P1$, $P2$, $C3$, and $C4$ sites. The reference and ground were attached with ECG disposable wet electrodes behind both subject's ears. The EEG signals were sampled at $256$Hz. The five male, right handed subjects participated voluntarily (ages $22-42$, mean $31$).
\begin{figure}[b]
\vspace{-0.3cm}
\begin{center}
	\includegraphics[width=0.8\linewidth]{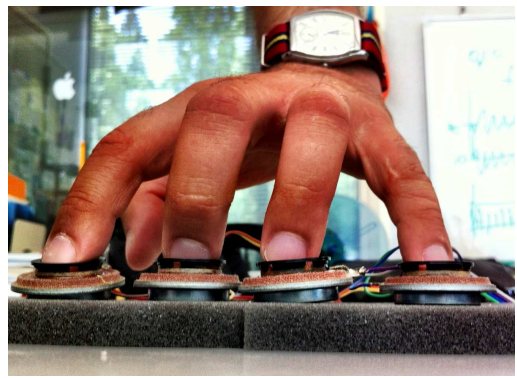}
\caption{The haptic BCI prototype with four tactile exciters positioned on a sponge to cancel a possible acoustic feedback.}
\label{fig:hand}
\end{center}
\end{figure}
For the EEG signal acquisition, stimulus triggering and response classification, an open source \texttt{BCI2000} \cite{bci2000book} system was chosen due to its flexibility and future ``no-cost'' application for patients in needs. \texttt{The BCI2000} system was set in P300 stimuli delivery and processing mode with the following parameters: 
\begin{enumerate}
	\item EEG high--pass and low--pass filters set to $0.1$Hz and $25$Hz respectively. Notch filter set for a $50$Hz frequency stopping.
	\item  stimulus length set to $10$ms delivered from \texttt{the Arduino UNO}; 
	\item stimulus onset asynchrony of $250$ms; 
	\item epoch length set to $800$ms; 
	\item epochs to average: $5$ to $8$ - subject performance dependent. 
\end{enumerate}
 
We tested LDA and Stepwise Linear Regression classifiers. The both resulted in similar outcomes. In this submission we report results with the latter one.

The stimulus delivery application was designed based on an open-source \texttt{Arduino UNO}~\cite{arduino} micro-controller connected via USB port to the host computer with a custom programmed communication software allowing vibrotactile stimuli delivery. We focus on the future portable and bedside application development.
The four tactile exciters (HiWave HIAX19C01-8; $19$mm; metal cup; $8\Omega$), as depicted in Figure~\ref{fig:hand}, were connected to digital outputs. The $10$ms square wave bursts with frequency of $1000$Hz were delivered as stimuli. The resulting square-wave-shape stimuli delivered ``the quick touch sensation'' for easier perception, as reported by the subjects. 

The messages from \texttt{the BCI2000} to the haptic application were sent via UDP port with triggers delivered to the custom patch developed in \texttt{a Max~6} suite~\cite{maxMSP}. The patch decoded \texttt{the BCI2000} events and sent the triggers to \texttt{the Arduino UNO} board using serial communication protocol. 

During the experiments subjects were instructed, as in classical P300 based oddball paradigms, to attend to a single finger tactile stimuli while ignoring the other three. Each finger had assigned a number $1-4$. Exactly the same number of vibrotactile bursts were delivered in a random order to each finger of the subject dominant hand and the subjects were instructed to choose (``spell'') the numbers as in usual ``copy spelling mode.''

\section{Results and Conclusions}

The results from our preliminary experiments with five healthy subjects are summarized in Table~\ref{tab:results}. The single subject was able to hit the maximum accuracy of $100\%$ while the remaining three were above a chance level of $25\%$. Testing the system with ALS patients is next on our agenda together with the improvement and optimization of the stimulation. The already obtained results have been comparable to~\cite{JNEtactileBCI2012}, or even better if we take into account the number of averages to reach the performance, and the number of electrodes used with our portable setting.

The reported possibility to combine open-source projects of \texttt{the Arduino UNO} and \texttt{the BCI2000}, together with compact EEG acquisition \texttt{g.MOBIlab+ \& g.SAHARA} hardware by \texttt{g.tec} is a step forward in development of the new portable, vision-free tactile BCI.
\begin{table}[h]
\caption{Peak interfacing accuracies (chance level $25\%$) and maximum bit--per--run--rate (BPRR)~\cite{aBCI2010} of  $17.14$~bit/min. The experiments conducted in the first session/day mode for subjects $\#~2-5$. The subject $\#~1$ was more experienced.}
\begin{center}
\begin{tabular}{| c | c | c | c |}
	\hline
	subject & max accuracy & BPRR [bit/min] \\ 
	\hline \hline 
	\#1 & $100\%$ & $17.14$ \\ 
	\#2 & $75\%$  & $7.92$ \\ 
	\#3 & $50\%$  & $1.56$ \\ 
	\#4 & $50\%$  &  $2.49$ \\ 
	\#5 & $75\%$  &  $5.94$  \\ 
	\hline
\end{tabular}
\end{center}
\vspace{-0.3cm}
\label{tab:results}
\end{table}
\small
\section*{Acknowledgments}
\small{This research was supported in part by a grant for ``Research and development of brain machine interface (BMI) using the acoustic spatial information for ALS patients'' within the scope of the Strategic Information and Communications R\&D Promotion Programme no.~$121803027$ of The Ministry of Internal Affairs and Communication in Japan.}

\end{document}